\PassOptionsToPackage{cmyk,table}{xcolor}
\documentclass[conference,flushend]{iaria} 
\pdfoutput=1 

\usepackage[babel=true,english=american]{csquotes}
\usepackage[USenglish]{babel}

\usepackage{array}
\usepackage{enumerate}
\usepackage{tikz}
\usepackage{caption}

\usepackage{biblatex}
\usepackage[%
	babel=true, 
	expansion=alltext,
	protrusion=alltext-nott, 
	nopatch=eqnum, 
	final 
]{microtype}

\newcolumntype{C}[1]{>{\centering\arraybackslash}p{#1}}

\title{Managing Forensic Recovery in the Cloud}

\author{
    \IEEEauthorblockN{George R.~S. Weir}
    \IEEEauthorblockA{%
        Department of Computer and Information Sciences\\
        University of Strathclyde\\
        Glasgow, UK\\
        e-mail: {\tt george.weir@strath.ac.uk}
    }
    \and
    \IEEEauthorblockN{Andreas Aßmuth\,\orcidlink{0009-0002-2081-2455} and Nicholas Jäger}
    \IEEEauthorblockA{%
    	University of Applied Sciences\\
    	OTH Amberg-Weiden\\
    	Germany\\
    	e-mail: {\tt$\lbrace$a.assmuth|n.jaeger$\rbrace$@oth-aw.de}
    }
}

\usepackage{ifpdf}
\ifpdf
\pdfoutput=1 
\pdftrue
\fi

\makeatletter
\def\ps@IEEEtitlepagestyle{
  \def\@oddfoot{\mycopyrightnotice}
  \def\@evenfoot{}
}
\def\mycopyrightnotice{
  {\footnotesize
    \begin{minipage}{0.8\textwidth}
    \centering
    Please cite as: \fullcite{selfref}.
    \end{minipage}
  }
}
\makeatother
\makeatletter
\let\blx@rerun@biber\relax
\makeatother
\usepackage[shortcuts]{extdash} 

\DeclareBibliographyCategory{selfref}
\addtocategory{selfref}{selfref}

\begin{document}
    
\maketitle

\begin{abstract}
As organisations move away from locally hosted computer services toward Cloud platforms, there is a corresponding need to ensure the forensic integrity of such instances. The primary reasons for concern are (i) the locus of responsibility, and (ii) the associated risk of legal sanction and financial penalty. Building upon previously proposed techniques for intrusion monitoring, we highlight the multi-level interpretation problem, propose enhanced monitoring of Cloud-based systems at diverse operational and data storage level as a basis for review of historical change across the hosted system and afford scope to identify any data impact from hostile action or ‘friendly fire’.
\end{abstract}

\begin{IEEEkeywords}
    \textbf{\textit{Cloud security; intrusion monitoring; message authentication codes; secret sharing.}}
\end{IEEEkeywords}

\section{Introduction}\label{sec:intro}
For many individuals, the primary use of Cloud computing is remote data storage. Presently, most major online Cloud service providers offer such storage. Apple users may engage iCloud as a supplement to local storage capacity and as an emergency backup for system configuration. Among similar
service offerings we find Google Drive, Microsoft OneDrive and Amazon Drive.\par 
Dropbox and its freemium business model, where users may register for a free account with a limited storage size and an option for more storage capacity and additional features for paid subscriptions, is also very popular. The broad appeal and immediate benefits from services of this type are apparent from the proliferation of such offerings, as underlined by the fact that many home broadband contracts include a measure of Cloud storage as standard. Thus, ``BT Cloud is a free service for BT Broadband customers that allows you to securely back up, access and share your precious files and folders''~\autocite{brittelecom}. Home broadband users will often rely on their remote storage and backup facility with little recognition that Cloud services are in operation.\par 
Despite the apparent speed with which consumers have adopted Cloud-based services, there is recognition that security issues can arise in the Cloud setting just as in the context of locally hosted systems~\autocite{nanavati,tirumala,zhou,chen}. When occasional security issues are reported in the media, the greatest concern may be the availability and privacy of their data~\autocite{bbc}.\par 
In the following, we outline the characteristics of risks that need to be accommodated in terms of forensic readiness. Firstly, we consider security risks arising from the network context, before focussing specifically on security issues in the Cloud setting. In section IV, we describe the concept of digital forensic readiness and, in Section V, explore the implications of applying this important aspect to the context of Cloud
services. We conclude by recommending greater attention to the requirements of Cloud forensic readiness, particularly with regard to the issue of multi-level interpretation. To this end, we propose enhanced monitoring of Cloud-based systems at diverse operational and data storage levels, as well as deployment of several previously advocated techniques for enhancing the security and resilience of recorded forensic readiness data.

\section{Network security risks}\label{sec:network-security-risks}
Addressing security risks is a familiar issue in the context of networked computing. In non-Cloud systems, the principal ingredients in management responses to security take three general forms:
\begin{itemize}
	\item System hardening
	\item Software defences
	\item Data backup
\end{itemize}
Firstly, system hardening is an attempt to render known threats ineffective. This includes ‘conventional’ measures that reduce vulnerability, such as authentication, identity management and access control~\autocite{takabi}, as well as acting to disable unnecessary services, applying regular software updates (patches) and gauging of the relevance and associated risks from newly published exploits~\autocite{carroll}. Modern Operating Systems have also been adapted to meet known cyber threats. Counter measures, like address space randomisation, mandatory access control or maybe sandboxing, are state of the art. In addition, advanced users might even build their own operating system and use selected kernel parameters to further harden their system. The second variety of response to address 
\begin{figure}[!hb]
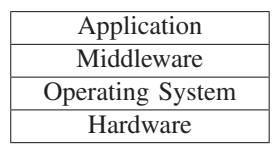

	\centering
	\begin{tabular}{|C{3cm}|}
		\hline
		Application\\
		\hline
		Middleware\vphantom{j}\\
		\hline
		Operating System\\
		\hline
		Hardware\vphantom{j}\\
		\hline
	\end{tabular}
	\caption{Layer-based model of a computing system}
	\label{fig:layer-based-model}
\end{figure}
security issues is the application of software defences. This ranges from antivirus provision to firewalls and may also include some variety of intrusion detection, usually rule-based~\autocite{ilgun} or anomaly-based~\autocite{garcia}.\par 
Any computing system may be described by a simple layer-based model as depicted in Figure~\ref{fig:layer-based-model}. Obviously, security on any higher layer strongly depends on access control mechanisms of lower layers. Even if users or service providers only aim for access control on a higher level to secure their application, these access control mechanisms in practice are more complex than those on lower layers. In addition, vulnerabilities or inadequate configuration on lower levels may lead to bypassing security measures on higher layers. Therefore, appropriate countermeasures are necessary on all layers.\par 
A third security measure is the provision of regular data backup, as a means of ensuring that any system failure or intrusion does not result in irretrievable data loss.

\section{Cloud security risks}\label{sec:cloud-security-risks}
Perhaps unsurprisingly, Cloud configurations are subject to levels of security risk that go beyond those affecting conventional networked computer systems. In consequence, the security measures outlined above may not be sufficient in the Cloud setting. In elaborating this claim, the Cloud issues are best illustrated with reference to the differing Cloud service offerings~\autocite{nistcloud}:
\begin{itemize}
	\item Infrastructure as a Service (IaaS);
	\item Platform as a Service (PaaS);
	\item Software as a Service (SaaS).
\end{itemize}
These models for Cloud service provision are helpfully elucidated by Gibson et. al.~\autocite{gibson}, as follows:
\begin{itemize}
	\item ``IaaS provides users with a web-based service that can be used to create, destroy, and manage virtual machines and storage. It can be used to meter the use of resources over a period of time, which in turn, can be billed back to users at a negotiated rate. It alleviates the users of the responsibility of managing the physical and virtualized infrastructure, while still retaining control over the operating system,
	configuration, and software running on the virtual machines'' [op. cit., p.~199].
	\item ``Platform-as-a-Service providers offer access to APIs, programming languages and development middleware which allows subscribers to develop custom applications without installing or configuring the development environment'' [op. cit., p.~200].
	\item ``Software-as-a-Service gives subscribed or pay-per-use users access to software or services that reside in the cloud and not on the user’s device'' [op. cit., p.~202].
\end{itemize}
Clearly, our earlier noted approaches to system security
are also applicable to Cloud-based systems. With an eye specifically on Cloud security, we can consider how each of
these service offerings may be at risk and what precautions
may be anticipated in response to these risks.

\subsection{Infrastructure as a Service}\label{subsec:iaas}
This kind of service seems most prone to the types of exploit that one would expect with conventional networked computers, principally, because, in most cases, such virtual machines will be presented to the Internet as networked hosts. Here, the customer is deploying a virtual machine with
associated Operating System and on-board software applications. This raises the prospect of vulnerabilities at network level, as well as application level issues, e.g., with Web systems and Database servers, Cross-Site Scripting (XSS) or SQL injections. Denial of service attacks are also a legitimate concern, especially since this kind of attack can achieve enormous bandwidths by using IoT devices for their purpose~\autocite{edwards}. For these reasons, \emph{system hardening} (especially, defending against known vulnerabilities) and \emph{software defences} are appropriate for IaaS, including precautions such as anti-malware, firewalls, and Intrusion Detection Systems. Provision of these features may be the responsibility of the Cloud Service Provider (CSP), who determines what OS and defensive capabilities are made available. In some settings, the customer may be in a position to bolster the native defences on the virtual system provided by the CSP.\par 
In similar vein, \emph{data backup} is likely to be required by the IaaS customer. Indeed, the protection of customer data may jointly be the concern of the customer and the CSP. The former may enable off-Cloud backup, to avoid a single source of failure. While the CSP may also offer data backup to a separate Cloud data storage facility.\par 
Despite reasonable expectation of such measures, there are indications that Cloud software infrastructure components are not always adequately secured from known vulnerabilities at the virtual machine level~\autocite{zhang}.

\subsection{Platform as a Service}\label{subsec:paas}
Computing facilities afforded to the customer of PaaS, are limited to the development of specific middleware or functional components. These services employ technologies such as Docker~\autocite{dhakate}, Containers~\autocite{pahl}, DevOps~\autocite{balalaie} and AWS Lambda~\autocite{villamizar}, in order to host customer-defined remote functionality. From a Cloud customer perspective, \emph{system hardening} seems to be irrelevant in this context in relation to the host Operating System. On the other hand, any code developed for use on the Cloud platform must be protected from illicit operations, e.g., process hijacking, output redirection or the elevation of privileges.\par 
\emph{Software defences} of the variety outlined above seem less relevant to the PaaS context since the operations supported by the middleware are limited to specific data processing and do not afford full operating system access or modification. The primary concern should be the operational effectiveness and resilience of the customer-defined operations. Clearly, such services may also be impaired through illicit access, e.g., stealing authentication details in order to alter code on the host system. Managing this area of concern lies primarily in the hands of the Cloud customer, with the assumption that the CSP will prevent unauthorised access to customer account details.

\subsection{Software as a Service}\label{subsec:saas}
SaaS provides the Cloud customer with remote access to third-party data processing facilities via micro-services~\autocite{namiot}, or RESTful services~\autocite{han}. Aside from network level attacks, such services should be protected from most other security concerns by having the host system hardened and equipped with suitable software defences. From the customer perspective, so long as their remote Cloud services operate effectively, without interruption or data loss, there would seem to be little cause for concern. Of course, the risk of aberrant customer-side behaviour may arise through social engineering exploits or disgruntled employee actions.\par 
This summary of security concerns affecting the three varieties of service has treated each Cloud model as an isolated networked computing facility. In reality, since the essence of Cloud provision is the virtualisation of services, our overview lacks one further important consideration, i.e., the possibility of service impairment as a result of activity at adjacent, upper or lower levels of the Cloud implementation.\par 
Clearly, any security aspects that affect the operational resilience of the underlying Cloud infrastructure is of direct concern to the CSP and can have a knock-on effect upon customer services. The underlying Cloud technology, i.e., the hardware and software configurations that provision our three Cloud models, may be subject to attack or deliberate manipulation in a fashion that impinges detrimentally upon the Cloud services supported by that particular hardware and software ensemble. This may be construed as a service attack ‘from below’. The scope for such attacks are precisely the characteristic exploits that may affect any networked host (listed earlier).\par
Attacks ‘from the side’ are a growing concern in Cloud security. ‘Side channel attacks’, originate with co-hosted customers who manipulate the behaviour of their virtual system to influence the behaviour of the host system and thereby affect co-hosted customers. Several studies suggest
that such ‘co-tenancy’, an essential feature of IaaS and PaaS, carries dangers. Thus, ``Physical co-residency with other tenants poses a particular risk''~\autocite{yzhang}, such as ``cache-based side-channel attacks''~\autocite{yzhang2}, and ``\emph{resource-freeing attacks} (RFAs)'' in which ``the goal is to modify the work-load of a victim VM in a way that frees up resources for the attacker’s VM''~\autocite{varadarajan}. Most worrying are contexts where one customer’s ‘malicious’ virtual machine seeks to extract information from another customer’s virtual machine on the same Cloud platform~\autocite{yzhang3}. Such risks to Cloud facilities are fundamental to their service provision.\par 
A final attack vector that threatens some Cloud systems is ‘from above’. In this case, poorly implemented virtual systems may afford scope for customers to ‘break free’ of their virtual system and access or directly affect the underlying Operating System or middleware/hypervisor. Clearly, it must be ensured that there is no information leakage from virtual machines and that attackers or malicious customers are not capable of
breaking out of the virtual machine and gaining access to the host OS or the virtual machines of other customers~\autocite{vateva}.\par 
The characteristics of these Cloud service offerings with associated security measures and the likely risk conditions are captured in Figure~\ref{fig:table}. The prospect of action from one Cloud user affecting another is described as intra-platform interference.

\section{Digital forensic readiness}\label{sec:digital-forensic-readiness}
The numbers of cases of network intrusion and data breach are on the rise: ``there is a massive increase in the records being compromised by external hacking~-- from roughly 49 million records in 2013 to 121 million and counting in 2015''~\autocite{secweek}.
\begin{figure}
	\centering
	\begin{tabular}{|C{1.92cm}|C{1.94cm}|C{1.54cm}|C{1.79cm}|}
		\hline
		\bf Service model & \bf Main features & \bf Security Measures & \bf Risks\\
		\hline
		Infrastructure (IaaS) & Virtual machines, Operating systems, Storage, Software applications & System hardening, Software defences, 		Data backup & Social engineering, Intrusion, Malware, Denial of service, Elevation of privileges, \emph{Intra-platform interference}\\
		\hline
		Platform (PaaS) & APIs, Programming languages, Development middleware, (Containers, Dockers, AWS Lambda, DevOps) & System hardening, Software defences & Social engineering, Elevation of privileges, \emph{Intra-platform interference}, \emph{Information leakage}\\
		\hline
		Software (SaaS) & Remote applications, Micro-services, RESTful services & System hardening, Software defences & Social engineering, \emph{Intra-platform interference}\\
		\hline
	\end{tabular}
	\caption{Summary of features, security measures and risks}
	\label{fig:table}
\end{figure}
One positive effect of this growth in unauthorized data access is the raised awareness of digital forensics (DF) and a marked change in its perception from a solely post-event reactive investigative tool to a pro-active policy to establish intelligence capabilities in advance of any incidents. This change in role reflects the concept of digital forensic readiness. Thus, ``Pro-active DF management must ensure that all business processes are structured in such a way that essential data and evidence will be retained to ensure successful DF investigations, should an incident occur''~\autocite[p.~18]{grobler}.\par 
One might define digital forensic readiness as ``having facilities in place to ensure the comprehensive capture and retention of all system event and user activity data that would be required post-incident in order to determine the precise nature of any data-loss, system modification or system impairment that results from intrusion, system misuse, or system failure''.\par 
Naturally, this concept of digital forensic readiness is equally applicable to Cloud systems and novel techniques have been proposed to facilitate the data collection that this entails~\autocite{kebande}. Yet, the Cloud context introduces particular problems with respect to forensic readiness.

\section{Cloud forensic recovery}\label{sec:cloud-forensic-recovery}
Forensic readiness in the Cloud is complicated by the variety of contexts in which Cloud services are deployed and the diversity of software settings in which security risks may arise. Forensic readiness must accommodate these complexities and, in turn, this suggests that a single
infrastructure-based digital forensic readiness solution may be infeasible.\par 
The primary reason for concern is the need to capture relevant data on system operation at the various operational levels of the Cloud system and any potential interaction across these levels. This means capturing program logs, system logs and user activity logs. In any end-customer Cloud facility, the data protected may not extend beyond any currently live information and data held in associated database systems. The ready recycle capability of Cloud services also has implications for the persistence of digital forensic evidence. An intrusion that steals data from a virtual machine and then seeks to reset that machine may well succeed in destroying evidence of the intrusion, thereby removing any forensic
traceability on the nature and quantity of stolen data.\par 
Neither is it sufficient to provide each distinct operational layer of Cloud systems with its own comprehensive forensic readiness. At best, this condition will allow for forensic data recovery for that operational layer. But there is no one-size-fits-all solution that can capture all state, interaction and performance data such as would ensure full Cloud forensic recovery. In fact, this insight reveals a fundamental problem that may impact upon Cloud forensic readiness.\par 
There are parallels here with issues in distributed systems and software architecture. Thus, ``distributed software systems are harder to debug than centralized systems due to the increased complexity and truly concurrent activity that is possible in these systems''~\autocite[p.~255]{bates}. Regardless of whether the Cloud setting is truly distributed in its realisation, its interconnected software functional layers represent a unique challenge when attempting to interpret the relationship between events or changes actioned at one functional level and the operational impact of such changes on other functional aspects of the services afforded by that Cloud.\par 
When considering Cloud systems, from the perspective of software architecture there may be an assumption of ‘a component- and message-based architectural style’ in which there is ‘a principle of limited visibility or substrate independence: a component within the hierarchy can only be
aware of components ``above'' it and is completely unaware of components which reside ``beneath'' it’~\autocite[p.~825]{medvidovic}.\par 
This multi-level interpretation problem is complicated by the fact that events considered anomalous at one level of service offering may arise through actions considered legitimate at a ‘lower’ level of software implementation. From the digital forensic readiness perspective, this underlines the requirement to go beyond capture of significant events across the Cloud service software and functional levels, since significance is an aspect that may cross the boundaries between such layers in the system as a whole. A hypothetical example may clarify this issue.\par 
A CSP customer may contract access to specific functional components (e.g., a Web service). The operational characteristics of the service are under the control of the CSP and not the customer. An authorised employee of the CSP may modify the algorithmic process and thereby affect the
outcome of any service use by the customer. While a change in operational behaviour of the service (i.e., an anomaly) may eventually be detected by the customer, there may be no anomalous activity evident at the level of CSP employee activity. The focus of subsequent forensic investigation may light initially on the nature of customer activity, since this is where the anomaly is apparent, but proper understanding of the issue requires that events across different functional levels of the Cloud system be apprehended.\par 
An informative view on this issue may be borrowed from Granular Computing~\autocite{yao}, which aims to develop computational models of complex systems, such as human intelligence. A key characteristic of this work is that it ‘stresses multiple views and multiple levels of understanding in each view’ [op. cit., p.~85]. Here, the emphasis is upon ‘holistic, unified views, in contrast to isolated, fragmented views. To achieve this, we need to consider multiple hierarchies and multiple levels in each hierarchy’ [op. cit., p.~88].\par 
Our proposal for adequate Cloud forensic readiness has two components. Firstly, there is a requirement for data capture. This is the obvious need to record any data at each layer of Cloud facility that may have a role to play in subsequent digital forensic analysis. Secondly, the captured
data must be stored off the system being monitored in a manner that both ensures the integrity of the logging and minimises the likelihood that the stored data can be compromised, either as a result of hostile action or ‘friendly fire’.\par 
To achieve adequate data capture, we require ‘state information’ and data management across differing levels of any Cloud service, from the lowest software level up to the most abstracted ‘user facing’ software component. On their own, such records will not be sufficient to fully capture the
potential interplay of differing software levels. For this purpose, subsequent digital forensic analytics will be required in order to establish a multi-dimensional representation of event chronology. This means that timestamps from events and data captured at different software levels of abstraction will be correlated to determine how events across the Cloud system are related.\par 
Our requirement for secure and resilient log storage can build upon default system logging that will be present within the Cloud implementation but this must be supplemented to achieve log reliability.\par 
Instead of using centralised log servers, which of course are attractive targets and easy to spot for attackers, we propose a different approach. In order to prevent adversaries from manipulating log files to hide their tracks, we use chained Message Authentication Codes (MACs) for each entry to the log file on each node. If state-of-the-art MACs are used, this makes it impossible to delete or manipulate text in the log files. Next, each node uses secret sharing techniques as proposed by Adi Shamir~\autocite{shamir} to divide the log file into parts. These parts are then sent to random other nodes which store these log data. Even if an adversary succeeds in taking over some of the nodes, he will need a certain number of these fragments to reconstruct the log data. But since for each log entry different nodes are chosen randomly as stated before, the attacker effectively needs to control the whole Cloud ecosystem to stay hidden. Further information on this solution can be found in our previous paper~\autocite{weir}.

\section{Conclusions}\label{sec:conclusions}
As organisations move increasingly away from locally hosted computer services toward Cloud-platforms, there is a corresponding need to ensure the forensic integrity of such instances. The primary reasons for concern are (i) the locus of responsibility, and (ii) the associated risk of legal sanction and financial penalty. In the first place, while Cloud service providers (CSPs) are responsible for the availability and robustness of their commercial offerings, they will not be responsible for the management of such services by their customers, nor for the data security associated with customer-level use of the Cloud services. Responsibility for these aspects resides with the CSP’s customers, whose data processing and data management are built upon the purchased Cloud services. In the second place, legislative demands on data protection, such as the forthcoming EU General Data Protection Regulation, will require companies to notify all breaches within 72 hours of discovery, or face significant
financial penalty.\par 
These concerns can be addressed and the business risk mitigated through development of forensic readiness in customer-level Cloud systems. We have argued that this requires a range of logging and data capture facilities across the Cloud system software infrastructure that maintain the possibility of tracking activity at different levels of software abstraction (the multi-level interpretation problem). Our second proposition is that such digital forensic readiness must be combined with techniques to ensure that logged data is incorruptible and robust. We have previously proposed techniques for intrusion monitoring that ensure log data credibility and provide robust decentralised log storage and recovery for post-hack scenarios~\autocite{weir}.

\renewcommand*{\bibfont}{\footnotesize}
\setlength{\labelnumberwidth}{0.45cm}
\printbibliography[notcategory=selfref]

\end{document}